\documentclass{research4cacm}

\usepackage{color}
\usepackage{url}
\usepackage{epstopdf}

\makeatletter
\def\@copyrightspace{\relax}
\makeatother

\begin{document}


\title{Deep Learning on FPGAs: Past, Present, and Future
%
}
%
%
%
%
%
\numberofauthors{3} 
%
\author{
%
%
\alignauthor
Griffin Lacey 
       \\
       \affaddr{University of Guelph}\\
       \affaddr{50 Stone Rd E}\\
       \affaddr{Guelph, Ontario}\\
       \email{laceyg@uoguelph.ca}
\alignauthor
Graham Taylor 
       \\
       \affaddr{University of Guelph}\\
       \affaddr{50 Stone Rd E}\\
       \affaddr{Guelph, Ontario}\\
       \email{gwtaylor@uoguelph.ca}
\alignauthor 
Shawki Areibi 
       \\
       \affaddr{University of Guelph}\\
       \affaddr{50 Stone Rd E}\\
       \affaddr{Guelph, Ontario}\\
       \email{sareibi@uoguelph.ca}
}

\maketitle
\begin{abstract}

The rapid growth of data size and accessibility in recent years has instigated a shift of philosophy in algorithm design for artificial intelligence. Instead of engineering algorithms by hand, the ability to learn composable systems automatically from massive amounts of data has led to ground-breaking performance in important domains such as computer vision, speech recognition, and natural language processing.  The most popular class of techniques used in these domains is called \textbf{\textit{deep learning}}, and is seeing significant attention from industry.  However, these models require incredible amounts of data and compute power to train, and are limited by the need for better hardware acceleration to accommodate scaling beyond current data and model sizes.  While the current solution has been to use clusters of graphics processing units (GPU) as general purpose processors (GPGPU), the use of field programmable gate arrays (FPGA) provide an interesting alternative.  Current trends in design tools for FPGAs have made them more compatible with the high-level software practices typically practiced in the deep learning community, making FPGAs more accessible to those who build and deploy models.  Since FPGA architectures are flexible, this could also allow researchers the ability to explore
model-level optimizations beyond what is possible on fixed architectures such as GPUs.  As well, FPGAs tend to provide high performance per watt of power consumption, which is of particular importance for application scientists interested in large scale server-based deployment or resource-limited embedded applications.  This review takes a look at deep learning and FPGAs from a hardware acceleration perspective, identifying trends and innovations that make these technologies a natural fit, and motivates a discussion on how FPGAs may best serve the needs of the deep learning community moving forward. 

\end{abstract}


\section{Introduction}

The effects of machine learning on our everyday life are far-reaching.  Whether you are clicking through personalized recommendations on websites, using speech to communicate with your smart-phone, or using face-detection to get the perfect picture on your digital camera, some form of artificial intelligence is involved.  This new wave of artificial intelligence is accompanied by a shift in philosophy for algorithm design.  Where past attempts at learning from data involved much ``feature engineering'' by hand using expert domain-specific knowledge, the ability to learn composable feature extraction systems automatically from massive amounts of example data has led to ground-breaking performance in important domains such as computer vision, speech recognition, and natural language processing.  The study of these data-driven techniques is called deep learning, and is seeing significant attention from two important groups of the technology community: researchers, who are interested in exploring and training these models to achieve top performance across tasks, and application scientists, who are interested in deploying these models for novel, real world applications.  However, both of these groups are limited by the need for better hardware acceleration to accommodate scaling beyond current data and algorithm sizes.

The current state of hardware acceleration for deep learning is largely dominated by using clusters of graphics processing units (GPU) as general purpose processors (GPGPU) \cite{coates2013deep}.  GPUs have orders of magnitude more computational cores compared to traditional general purpose processors (GPP), and allow a greater ability to perform parallel computations.  In particular, the NVIDIA CUDA platform for GPGPU programming is most dominant, with major deep learning tools utilizing this platform to access GPU acceleration \cite{chetlur2014cudnn,jia2014caffe,bergstra2010theano,collobert2002torch}.  More recently, the open parallel programming standard OpenCL has gained traction as an alternative tool for heterogeneous hardware programming, with interest from these popular tools gaining momentum.  OpenCL, while trailing CUDA in terms of support in the deep learning community, has two unique features which distinguish itself from CUDA.  First is the open source, royalty-free standard for development, as opposed to the single vendor support of CUDA.  The second is the support for a wide variety of alternative hardware including GPUs, GPPs, field programmable gate-arrays (FPGA), and digital signal processors (DSP).    

\subsection{The Case for FPGAs}

The imminent support for alternative hardware is especially important for FPGAs, a strong competitor to GPUs for algorithm acceleration.  Unlike GPUs, these devices have a flexible hardware configuration, and often provide better performance per watt than GPUs for subroutines important to deep learning, such as sliding-windows computation \cite{fowers2012performance}.  However, programming of these devices requires hardware specific knowledge that many researchers and application scientists may not possess, and as such, FPGAs have been often considered a specialist architecture.  Recently, FPGA tools have adopted software-level programming models, including OpenCL, which has made them a more attractive option for users trained in mainstream software development practices. 

For researchers considering a variety of design tools, the selection criteria is typically related to having user-friendly software development tools, flexible and upgradeable ways to design models, and fast computation to reduce the training time of large models.  Deep learning researchers will benefit from the use of FPGAs given the trend of higher abstraction design tools which are making FPGAs easier to program, the reconfigurability which allows customized architectures, and the large degree of parallelism which will accelerate execution speeds.  

For application scientists, while similar tool level preferences exist, the emphasis for hardware selection is to maximize performance per watt of power consumption, reducing costs for large scale operations.  Deep learning application scientists will benefit from the use of FPGAs given the strong performance per watt that typically accompanies the ability to customize the architecture for a particular application. 

FPGAs serve as a logical design choice which appeal to the needs of these two important audiences.  This review takes a look at the current state of deep learning on FPGAs, as well as current developments which serve to bridge these two technologies.  As such, this review serves three important purposes.  First, it identifies the opportunity that exists within the deep learning community for exploring new hardware acceleration platforms, and shows FPGAs as an ideal choice.  Next, it outlines the current state of FPGA support for deep learning, identifying potential limitations.  Finally, it makes key recommendations of future directions for FPGA hardware acceleration that would help in solving the deep learning problems of tomorrow.

\section{Deep Learning}

Conventional approaches to artificial intelligence focused on using computation to solve problems analytically, requiring explicit knowledge about a given domain \cite{bengio2011expressive}.  For simple problems this approach was adequate, as the programs engineered by hand were small, and domain experts could carefully transform the modest amount of raw data into useful representations for learning.  However, advances in artificial intelligence created interest in solving more complex problems, where knowledge is not easily expressed explicitly.  Expert knowledge about problems such as face recognition, speech transcription, and medical diagnosis is difficult to express formally, and conventional approaches to artificial intelligence failed to account for the implicit information stored in the raw data.  Moreover, tremendous growth in data acquisition and storage means that using this implicit information is more important than ever.  Recently, these types of applications are seeing state-of-the-art performance from a class of techniques called deep learning, where this implicit information is discovered automatically by learning task-relevant features from raw data.  Interest in this research area has led to several recent reviews \cite{lecun2015nature,Schmidhuber,bengio2009learning}.  

The field of deep learning emerged around 2006 after a long period of relative disinterest around neural networks research. Interestingly, the early successes in the field were due to unsupervised learning-- techniques that can learn from unlabeled data. Specifically, unsupervised learning was used to ``pre-train'' (initialize) the layers of deep neural networks, which were thought at the time to be too difficult to train with the usual methods, i.e.~gradient backpropagation. However, with the introduction of GPGPU computing and the availability of larger datasets towards the end of the 2000's and into the current decade, focus has shifted almost exclusively to supervised learning. In particular, there are two types of neural network architectures that have received most of the attention both in research and industry. These are multi-layer perceptrons (MLP) and convolutional neural networks (CNN). Essentially all of the research on FPGA-based deep learning has focused on one of these architectures, and therefore we briefly describe them below.

Before describing any specific architecture, however, it is worth noting several characteristics of most deep learning models and applications that, in general, make them well-suited for parallellization using hardware accelerators.

\noindent {\bf Data parallelism} -- The parallelism inherent in pixel-based sensory input (e.g.~images and video) manifests itself in operations that apply concurrently to all pixels or local regions. As well, the most popular way of training models is not by presenting it with a single example at a time, but by processing ``minibatches'' of typically hundreds or thousands of examples. Each example in a minibatch can be processed independently.\\
\noindent {\bf Model parallelism} -- These biologically-inspired models are composed of redundant processing units which can be distributed in hardware and updated in parallel. Recent work on accelerating CNNs using multiple GPUs has used very sophisticated strategies to balance data and model-based parallism such that different parts of the architecture are parallelized in different, but optimal ways \cite{krizhevsky2014one}.\\
\noindent {\bf Pipeline Parallelism} -- The feed-forward nature of computation in architectures like MLPs and CNNs means that hardware which is well suited to exploit pipeline parallelism (e.g.~FPGAs) can offer a particular advantage.  While GPPs and GPUs rely on executing parallel threads on multiple cores, FPGAs can create customized hardware circuits which are deeply pipelined and inherently multithreaded. 

\subsection{Multi-layer Perceptrons}

\begin{figure*}[!htb]
\centering
\includegraphics[width=0.88\textwidth]{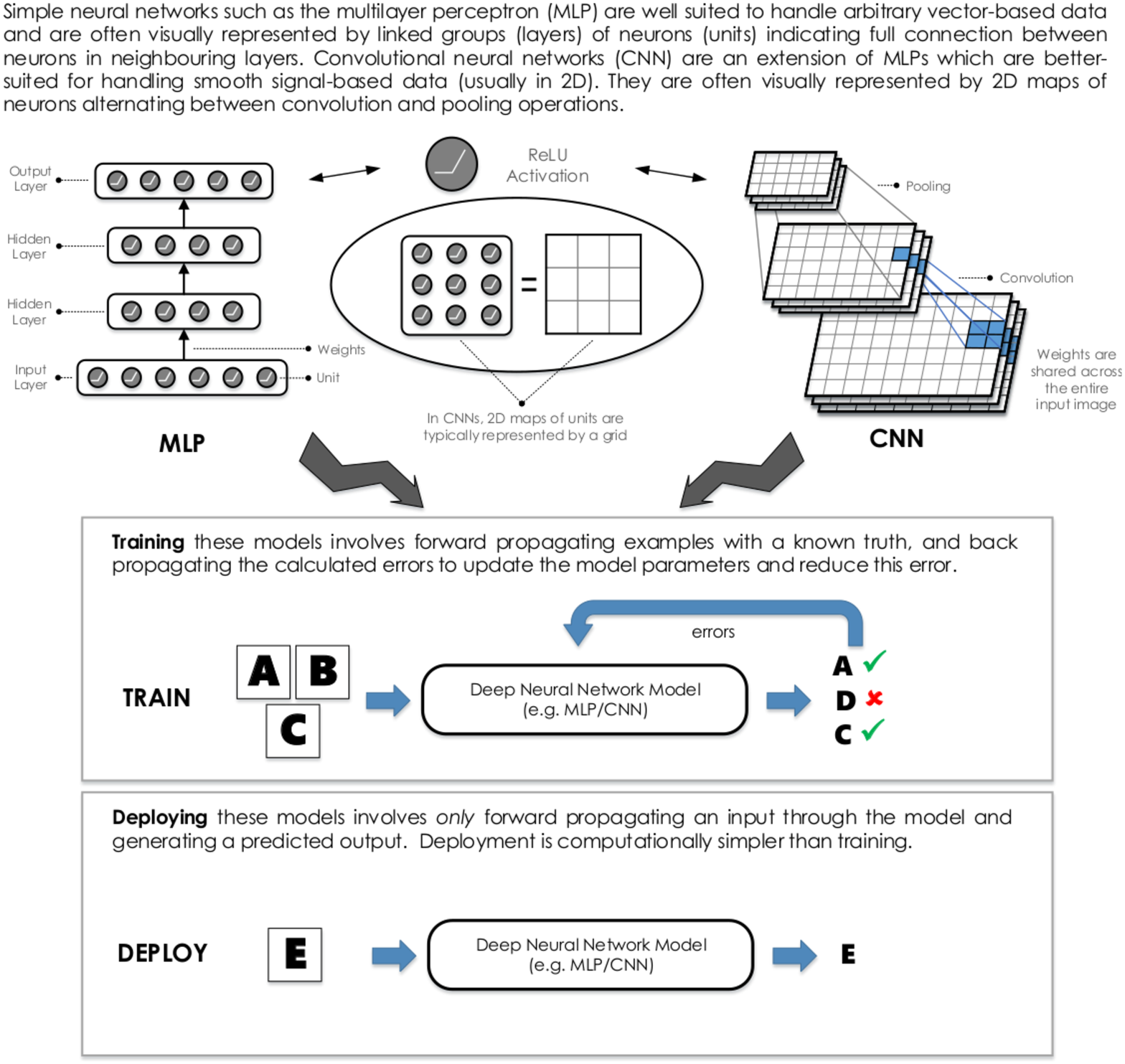}
\caption{Differences and similarities between MLPs and CNNs.}
\label{fig:NNvsCNN}
\end{figure*}

Simple feed-forward deep networks are known as multi-layer perceptrons (MLP), and are the backbone of deep learning \cite{Bengio-et-al-2015-Book}.  To describe these models using neural network terminology, we refer to the examples fed to these models as \textit{inputs}, the predictions produced from these models as \textit{outputs}, each modular sub-function as a \textit{layer} with \textit{hidden layers} referring to those layers between the first (input) layer and last (output) layer, each scalar output of one of these layers as a \textit{unit} (analogous to a neuron in the biological inspiration of these models), and each connection between units as a \textit{weight} (analogous to a synapse), which define the function of the model as they are the parameters that are adjusted during training \cite{bengio2009learning}. Collections of units are sometimes referred to as \textit{features}, as to draw similarities to the traditional idea of features in conventional machine learning, which were designed by domain experts. To prevent the entire network from collapsing to a linear transformation, each unit applies an element-wise nonlinear operation to its input, with the most popular choice being the rectified linear unit (ReLU).  A basic MLP is illustrated in Figure \ref{fig:NNvsCNN}. 

\subsection{Convolutional Neural Networks}

Deep convolutional neural networks (CNN) are currently the most popular deep learning architectures, especially for pixel-based visual recognition tasks.  More formally, these networks are designed for data that has a measure of spatial or temporal continuity.  This inspiration is drawn largely on work from Hubel and Wiesel, who described the function of a cat's visual cortex as being sensitive to small sub-regions of the visual field \cite{hubel1968receptive}.  Commonly, spatial continuity in data is found in images where a pixel at location $(i,j)$ shares similar intensity or color properties to its neighbours in a local region of the image.  

CNNs are composed of various combinations of a few important layer types.  These layers, in comparison to MLPs, are constructed as a 2D arrangement of units called \textit{feature maps}.  Convolution layers, analogous to the linear feature extraction operation of MLPs, are parameterized by learnable filters (kernels), which have local connections to a small receptive field of the input feature map and shared at all locations of the input. Feature extraction in a CNN amounts to convolution with these filters.  Pooling layers apply a simple reduction operation (e.g.~a max or average) to local regions of the feature maps. This reduces the size of the feature maps, which is favorable to computation and reducing parameters, but also yields a small amount of shift-invariance.
Finally, in recognition applications CNNs typically apply one or more fully connected layers (the same layers used in MLPs) towards the output layer in order to reduce the spatially and/or temporally organized information in feature maps to a decision, such as a classification or regression.

\section{FPGA\MakeLowercase{s}}

Traditionally, when evaluating hardware platforms for acceleration, one must inevitably consider the trade-off between flexibility and performance.  On one end of the spectrum, general purpose processors (GPP) provide a high degree of flexibility and ease of use, but perform relatively inefficiently.  These platforms tend to be more readily accessible, can be produced cheaply, and are appropriate for a wide variety of uses and reuses.  On the other end of the spectrum, application specific integrated circuits (ASIC) provide high performance at the cost of being inflexible and more difficult to produce.  These circuits are dedicated to a specific application, and are expensive and time consuming to produce.

FPGAs serve as a compromise between these two extremes.  They belong to a more general class of programmable logic devices (PLD) and are, in the most simple sense, a reconfigurable integrated circuit.  As such, they provide the performance benefits of integrated circuits, with the reconfigurable flexibility of GPPs. At a low-level, FPGAs can implement sequential logic through the use of flip-flops (FF) and combinational logic through the use of look-up tables (LUT).  Modern FPGAs also contain hardened components for commonly used functions such as full processor cores, communication cores, arithmetic cores, and block RAM (BRAM).  In addition, current FPGA trends are tending toward a system-on-chip (SoC) design approach, where ARM coprocessors and FPGAs are commonly found on the same fabric.  The current FPGA market is dominated by Xilinx and Altera, accounting for a combined 85 percent market share \cite{bacon2013fpga}.  In addition, FPGAs are rapidly replacing ASICs and application specific standard products (ASSP) for fixed function logic.  The FPGA market is expected to reach the \$10 billion mark by 2016 \cite{bacon2013fpga}.  

For deep learning, FPGAs provide an obvious potential for acceleration above and beyond what is possible on traditional GPPs.  Software-level execution on GPPs rely on the traditional Von Neumann architecture, which stores instructions and data in external memory to be fetched when needed.  This is the motivation for caches, which alleviate much of the expensive external memory operations \cite{bacon2013fpga}.  The bottleneck in this architecture is the processor and memory communication, which severely cripples GPP performance, especially for the memory-bound techniques frequently required in deep learning.  In comparison, the programmable logic cells on FPGAs can be used to implement the data and control path found in common logic functions, which do not rely on the Von Neumann architecture.  They are also capable of exploiting distributed on-chip memory, as well as large degrees of pipeline parallelism, which fit naturally with the feed-forward nature deep learning methods.  Modern FPGAs also support partial dynamic reconfiguration, where part of the FPGA can be reprogrammed while another part of the FPGA is being used.  This can have implications for large deep learning models, where individual layers could be reconfigured on the FPGA while not disrupting ongoing computation in other layers.  This would accommodate models which may be too large to fit on a single FPGA, and also alleviate expensive global memory reads by keeping intermediate results in local memory.    

Most importantly, when compared to GPUs, FPGAs offer a different perspective on what it means to accelerate designs on hardware.  With GPUs and other fixed architectures, a software execution model is followed, and structured around executing tasks in parallel on independent compute units.  As such, the goal in developing deep learning techniques for GPUs is to adapt algorithms to follow this model, where computation is done in parallel, and data interdependence is ensured.  In contrast, FPGA architecture is tailored for the application.  When developing deep learning techniques for FPGAs, there is less emphasis on adapting algorithms for a fixed computational structure, allowing more freedom to explore algorithm level optimizations.  Techniques which require many complex low-level hardware control operations which cannot be easily implemented in high-level software languages are especially attractive for FPGA implementations.  However, this flexibility comes at the cost of large compile (place and route) times, which is often problematic for researchers who need to quickly iterate through design cycles.  

In addition to compile time, the problem of attracting researchers and application scientists, who tend to favour high-level programming languages, to develop for FPGAs has been especially difficult.  While it is often the case that being fluent in one software language means one can easily learn another, the same cannot be said for translating skills to hardware languages.  The most popular languages for FPGAs have been Verilog and VHDL, both examples of hardware description languages (HDL).  The main difference between these languages and traditional software languages, is that HDL is simply describing hardware, whereas software languages such as C are describing sequential instructions with no need to understand hardware level implementation details.  Describing hardware efficiently requires a working knowledge of digital design and circuits, and while some of the low level implementation decisions can be left to automatic synthesizer tools, this does not always result in efficient designs.  As a result, researchers and application scientists tend to opt for a software design experience, which has matured to support a large assortment of abstractions and conveniences that increase the productivity of programmers.  These trends have pushed the FPGA community to now favour design tools with a high-level of abstraction.

\subsection{High-Level Abstraction Tools}

Both Xilinx and Altera have favoured the use of high-level design tools which abstract away many of the challenges of low level hardware programming.  These tools are commonly termed high-level synthesis (HLS) tools, which translate high-level designs into low-level register-transfer level (RTL) or HDL code.  A good overview of HLS tools is presented in \cite{bacon2013fpga}, where they are grouped into five main categories:  model-based frameworks, high-level language based frameworks, HDL-like languages, C-based frameworks, and parallel computing frameworks (i.e.~CUDA/OpenCL).  While it is important to understand these different types of abstraction tools, this review focuses on parallel computing frameworks, as they provide the most sensible path to join deep learning and FPGAs. 

\begin{figure*}[!htbp]
\centering
\includegraphics[width=0.88\textwidth]{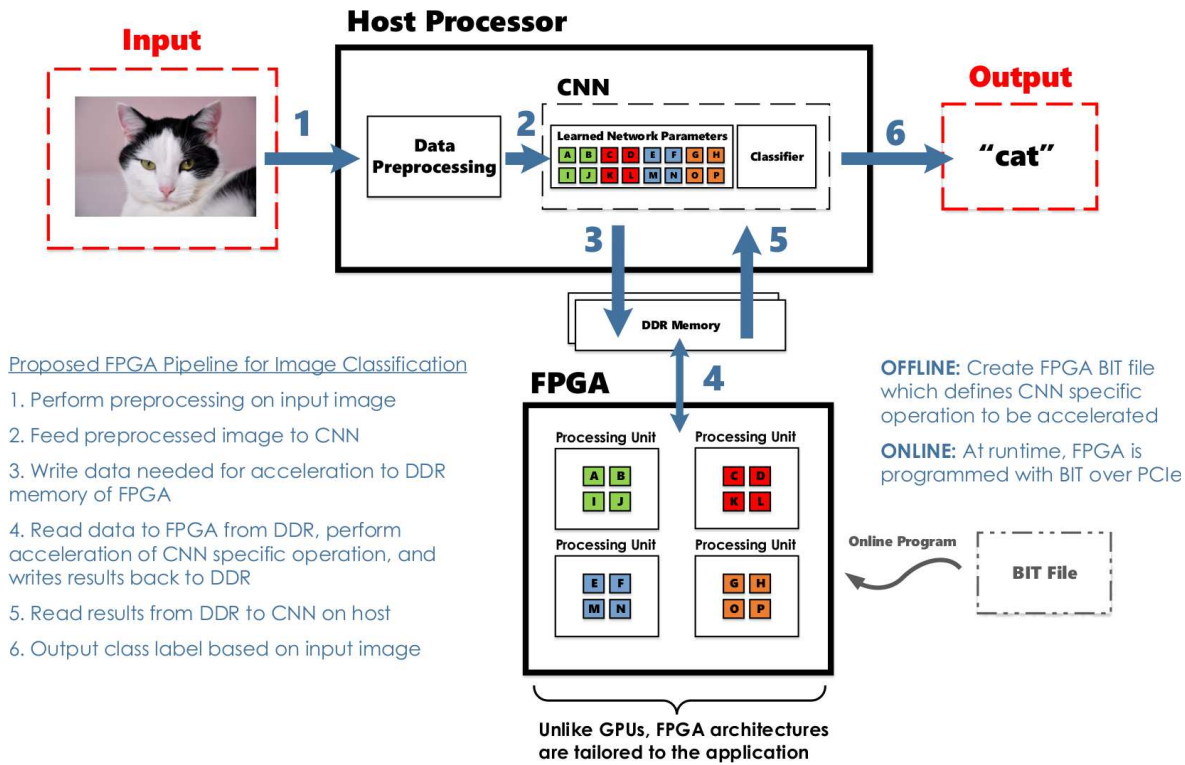}
\caption{Proposed deployment flow for image classification using FPGA for acceleration.}
\label{fig:workflow}
\end{figure*}      

\subsection{OpenCL}

OpenCL is an open source, standardized framework for algorithm acceleration on heterogeneous architectures.  As a C-based language (C99), programs written in OpenCL can be executed transparently on GPPs, GPUs, DSPs, and FPGAs.  Similar to CUDA, OpenCL provides a standard framework for parallel programming, as well as low-level access to hardware.  While both CUDA and OpenCL provide similar functionality to programmers, key differences between them have left most people divided.  Since CUDA is the current choice for most popular deep learning tools, it is important to discuss these differences in detail, in the interest of demonstrating how OpenCL could be used for deep learning moving forward.

The major difference between OpenCL and CUDA is in terms of ownership.  CUDA is a proprietary framework created by the hardware manufacturer NVIDIA, known for manufacturing high performance GPUs.  OpenCL is open-source, royality-free, and is maintained by the Khronos group.  This gives OpenCL a unique capability compared to CUDA: OpenCL can support programming a wide variety of hardware platforms, including FPGAs.  However, this flexibility comes at a cost, where all supported platforms are not guaranteed to support all OpenCL functions.  In the case of FPGAs, only a subset of OpenCL functions are currently supported.  While a detailed comparison of OpenCL and CUDA is outside the scope of this paper, performance of both frameworks has been shown to be very similar for given applications \cite{fang2011comprehensive}.
        
Beginning in late 2013, both Altera and Xilinx started to adopt OpenCL SDKs for their FPGAs \cite{czajkowski2012opencl,sdaccel}.  This move allowed a much wider audience of software developers to take advantage of the high-performance and low power benefits that come with designing for FPGAs.  Conveniently, both companies have taken similar approaches to adopting OpenCL for their devices. 

\subsection{Proposed Design Flow for Deep Learning Development}

To successfully integrate FPGAs into deep learning design flows, the needs of researchers and application scientists, who are familiar with GPU design flows, need to be considered.  While this is a challenge given the architectural differences of FPGAs and GPUs, we believe this goal is achievable.  

The main challenge is related to design compile time. Both Altera and Xilinx support primarily offline compiling for OpenCL kernels.  The main reason for this is that OpenCL kernel compilation time for both vendors is on the order of tens of minutes to hours, whereas compiling generic OpenCL kernels for GPPs/GPUs is on the order of milliseconds to seconds.  Obviously, this makes iterating through the design phase challenging if the compilation time is hours for each design iteration.  However, this is not necessarily futile for deep learning, as deep learning tools often reuse the same pre-compiled kernels during the design phase.  As well, deep learning tools supported by CUDA use a similar methodology, as CUDA employs a just-in-time compiling approach.  As such, doing one-time offline compilation of commonly used deep learning kernels is a reasonable compromise that does not limit application scientists who are usually not designing these kernels, but just interested in using them.  Figure \ref{fig:workflow} shows an example flow for deployment of an image classification model.

\begin{figure*}[!htbp]
\centering
\includegraphics[width=0.88\textwidth]{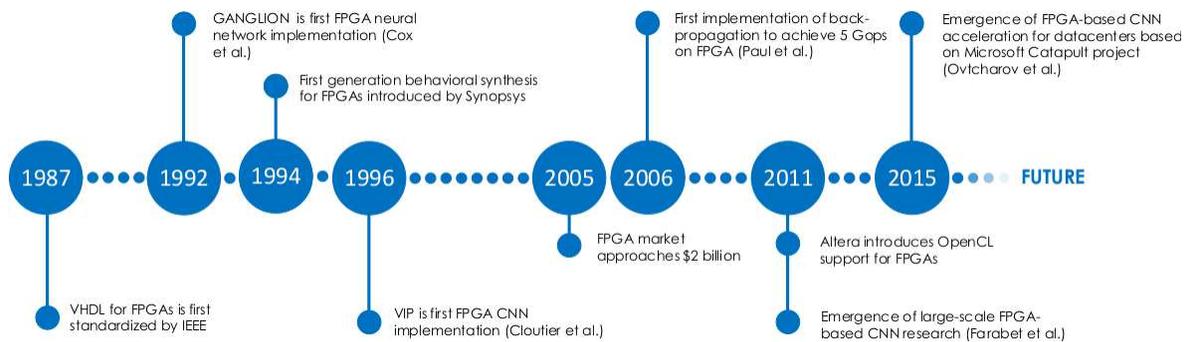}
\caption{Timeline of important events in FPGA deep learning research.}
\label{fig:timeline}
\end{figure*}

For researchers who do have an interest in kernel design, both Altera and Xilinx support integrated kernel profiling, debugging, and optimization in their OpenCL tools.  These features give researchers the ability to speed up and optimize the development of kernels.  As well, both companies have support for kernel simulation in software, which can circumvent the hassle of dealing with long compile times when debugging non-hardware issues such as semantic or syntax errors.

\section{A Review of CNNs on FPGAs}

One of the most limiting hardware realizations for deep learning techniques on FPGAs is design size.  The tradeoff between design reconfigurability and density means that FPGA circuits are often considerably less dense than hardware alternatives, and so implementing large neural networks has not always been possible.  However, as modern FPGAs continue to exploit smaller feature sizes to increase density, and incorporate hardened computational units alongside generic FPGA fabric, deep networks have started being implemented on single FPGA systems.  A brief timeline of important events in FPGA deep learning research is seen in Figure \ref{fig:timeline}.    

The first FPGA implementations of neural networks began appearing in the early 1990's, with the first implementation credited to Cox et al. in 1992 \cite{cox1992ganglion}.  However, the first FPGA implementations of CNNs began appearing a few years later.  Cloutier et al.~were among the first to explore these efforts, but were strongly limited by FPGA size constraints at the time, leading to the use of low-precision arithmetic \cite{cloutier1996vip}.  In addition, because FPGAs at this time did not contain the dense hardened multiply-accumulate (MAC) units that are present in today's FPGAs, arithmetic was also very slow in addition to being resource expensive.  Since this time, FPGA technology has changed significantly.  Most notably, there has been a large increase in the density of FPGA fabric, motivated by the decreasing feature (transistor) size, as well as an increase in the number of hardened computational units present in FPGAs.  State-of-the-art FPGA implementations of CNNs take advantage of both of these design improvements.

To the best of our knowledge, state-of-the-art performance for forward propagation of CNNs on FPGAs was achieved by a team at Microsoft.  Ovtcharov et al.~have reported a throughput of 134 images/second on the ImageNet 1K dataset \cite{krizhevsky2012imagenet}, which amounts to roughly 3x the throughput of the next closest competitor, while operating at {25~W} on a Stratix V D5 \cite{ovtcharov2015accelerating}.  This performance is projected to increase by using top-of-the-line FPGAs, with an estimated throughput of roughly 233 images/second while consuming roughly the same power on an Arria 10 GX1150.  This is compared to high-performing GPU implementations (Caffe + cuDNN), which achieve 500-824 images/second, while consuming 235~W.  Interestingly, this was achieved using Microsoft-designed FPGA boards and servers, an experimental project which integrates FPGAs into datacenter applications.  This project has claimed to improve large-scale search engine performance by a factor of two, showing promise for this type of FPGA application \cite{catapult}. 

Other strong efforts include the design proposed by Zhang et al., referenced above as the closest competitor achieving a throughput of 46 images/second on a Virtex 7 485T, with an unreported power consumption \cite{zhang2015optimizing}.  In this paper, Zhang et al. show their work to outperform most of the main strong competitors in this field, including \cite{cadambi2010programmable,chakradhar2010dynamically,farabet2009cnp,peemen2013memory,sankaradas2009massively}.  Most of these implementations contain architecturally similar designs, commonly using off-chip memory access, configurable software layers, buffered input and output, and many parallel processing elements implemented on FPGA fabric (commonly used to perform convolution).  However, important FPGA specific differences exist, such as using different memory sub-systems, data transfer mechanisms, soft-cores, LUT types, operation frequencies, and entirely different FPGAs, to name a few.  As a result, it is hard to determine specific optimal architecture decisions, as more research is needed.    

Since pre-trained CNNs are algorithmically simple and computationally efficient, most FPGA efforts have involved accelerating the forward propagation of these models, and reporting on the achieved throughput.  This is often of most importance to application engineers who wish to use pre-trained networks to process large amounts of data as quickly and efficiently as possible.  However, this only represents one aspect of CNN design considerations for FPGAs, as accelerating backward propagation on FPGAs is also an area of interest.  Paul et al.~were the first to completely parallelize the learning phase on a Virtex E FPGA in 2006 \cite{paul2006back}.

\begin{table*}[!htbp]
\centering
\caption{Overview of Deep Learning Frameworks with OpenCL Support}
\begin{tabular}{|c|l|l|l|c|} 
\hline
\multicolumn{1}{|c|}{\textbf{Tool}} & \multicolumn{1}{c|}{\textbf{Core Language}} & \multicolumn{1}{c|}{\textbf{Bindings}} & \multicolumn{1}{c|}{\textbf{OpenCL}} & \multicolumn{1}{c|}{\textbf{User Base}}
\\ \hline
\textit{Caffe} & C++ & \begin{tabular}[c]{@{}l@{}}Python\\ MATLAB\end{tabular} & Partial Support & Large 
\\ \hline
\textit{Torch} & Lua & \multicolumn{1}{c|}{-} & Partial Support & Large
\\ \hline
\textit{Theano} & Python & \multicolumn{1}{c|}{-} & Minimal Support & Large
\\ \hline
\textit{DeepCL} & C++ & \begin{tabular}[c]{@{}l@{}}Python\\ Lua\end{tabular} & Full Support & Moderate
\\ \hline
\end{tabular}
\label{tab:ocl}
\end{table*}  

\section{Looking Forward}

The future of deep learning on FPGAs, and in general, is largely dependant on scalability.  For these techniques to succeed on the problems of tomorrow, they must scale to accommodate data sizes and architectures that continue to grow at an incredible rate.  FPGA technology is adapting to support this trend, as the hardware is headed toward larger memory, smaller feature sizes, and interconnect improvements to accommodate multi-FPGA configurations.  The Intel acquisition of Altera, along with the partnership of IBM and Xilinx, indicate a change in the FPGA landscape which may also see the integration of FPGAs in consumer and data center applications in the very near future.  In addition, design tools will likely tend toward higher levels of abstraction and software-like experiences, in an effort to attract a wider technical range of users.   

\subsection{Popular Deep Learning Software Tools}

Of the most popular software packages for deep learning, several have began to take notice of the need for OpenCL support in addition to CUDA support.  This support will make FPGAs more accessible for the purposes of deep learning.  While, to our knowledge, no deep learning tools exist yet which explicitly support FPGAs, the following list (summarized in Table \ref{tab:ocl}) details some of the notable OpenCL efforts which move these tools in that direction:

\begin{itemize}
\item Caffe, developed by the Berkeley Vision and Learning Center, has unofficial support for OpenCL under the name project GreenTea \cite{greentea}.  There is also an AMD version of Caffe that supports OpenCL \cite{amd}.

\item Torch, a scientific computing framework written in Lua, is widely used and has unofficial support for OpenCL under the project CLTorch \cite{cltorch}. 

\item Theano, developed by the University of Montreal, has unofficial support for OpenCL under the work-in-progress gpuarray backend \cite{gpuarray}.

\item DeepCL is an OpenCL library to train deep convolutional neural networks, developed by Hugh Perkins \cite{deepcl}.
\end{itemize}

For those new to this field who are looking to choose between these tools, our recommendation is to start with Caffe due to its popularity, support, and easy to use interface.  As well, using Caffe's ``model zoo'' repository, it is easy to experiment with popular pre-trained models.

\subsection{Increasing Degrees of Freedom for Training}

While one may expect the process of training machine learning algorithms to be fully autonomous, in practice there are tunable hyper-parameters that need to be adjusted.  This is especially true for deep learning, where complexity of the model in terms of number of parameters is often accompanied by many possible combinations of hyper-parameters.  The number of training iterations, the learning rate, mini-batch size, number of hidden units, and number of layers are all examples of hyper-parameters that can be adjusted.  The act of tuning these values is equivalent to selecting which model, among the set of all possible models, is best for a particular problem. Traditionally, hyper-parameters have been set by experience or systematically by grid search or more effectively, random search \cite{bergstra2012random}. Very recently, researchers have turned to adaptive methods, which exploit the results of hyper-parameter attempts. Among these, Bayesian Optimization \cite{snoek2012practical} is the most popular.

  Regardless of the method selected to tune hyper-parameters, current training procedures using fixed architectures are somewhat limited in their ability to grow these sets of possible models, meaning that we may be viewing the solution space through a very narrow lens.  Fixed architectures make it much easier to explore hyper-parameter settings within models (e.g.~number of hidden units, number of layers) but difficult to explore settings between models (i.e.~different \emph{types} of models) as the training of models which do not conveniently conform to a particular fixed architecture may be very slow.  The flexible architecture of FPGAs, however, may be better suited for these types of optimizations, as a completely different hardware structure can be programmed and accelerated at runtime.

\subsection{Low Power Compute Clusters}

One of the most intriguing aspects of deep learning models is the ability to scale.  Whether the purpose is to discover complex high level features in data, or to increase performance for data center applications, deep learning techniques are often scaled up across multi-node computing infrastructures.  Current solutions to this problem involve using clusters of GPUs with Infiniband interconnects and MPI to allow high levels of parallel computing power and fast data transfer between nodes \cite{coates2013deep}.  However, as the workloads of these large scale applications become increasingly heterogeneous, the use of FPGAs may prove to be a superior alternative.  The programmability of FPGAs would allow reconfigurability based on the application and workload, and FPGAs provide an attractive performance/watt that would lower costs for the next generation of data centers.

\section{Conclusion}

When addressing the hardware needs of deep learning, FPGAs provide an attractive alternative to GPUs and GPPs.  In particular, the ability to exploit pipeline parallelism and achieve an efficient rate of power consumption give FPGAs a unique advantage over GPUs and GPPs for common deep learning practices.  As well, design tools have matured to a point where integrating FPGAs into popular deep learning frameworks is now possible.  Looking forward, FPGAs can effectively accommodate the trends of deep learning and provide architectural freedom for exploration and research. 

\bibliographystyle{abbrv} 
\bibliography{bib}  

\begin{thebibliography}{10}

\bibitem{amd}
{Caffe-OpenCL}.
\newblock \url{https://github.com/amd/OpenCL-caffe/wiki}, 2015.

\bibitem{greentea}
{Caffe:} project greentea.
\newblock \url{https://github.com/BVLC/caffe/pull/2195}, 2015.

\bibitem{deepcl}
{DeepCL}.
\newblock \url{https://github.com/hughperkins/DeepCL}, 2015.

\bibitem{catapult}
{Microsoft Research:} catapult, 2015.

\bibitem{gpuarray}
{Theano:} gpuarray backend.
\newblock \url{http://deeplearning.net/software/libgpuarray/index.html}, 2015.

\bibitem{cltorch}
{Torch:} cltorch.
\newblock \url{https://github.com/hughperkins/cltorch}, 2015.

\bibitem{sdaccel}
{Xilinx:} sdaccel, 2015.

\bibitem{bacon2013fpga}
D.~F. Bacon, R.~Rabbah, and S.~Shukla.
\newblock Fpga programming for the masses.
\newblock {\em Communications of the ACM}, 56(4):56--63, 2013.

\bibitem{bengio2009learning}
Y.~Bengio.
\newblock Learning deep architectures for ai.
\newblock {\em Foundations and trends{\textregistered} in Machine Learning},
  2(1):1--127, 2009.

\bibitem{bengio2011expressive}
Y.~Bengio and O.~Delalleau.
\newblock On the expressive power of deep architectures.
\newblock In {\em Algorithmic Learning Theory}, pages 18--36. Springer, 2011.

\bibitem{Bengio-et-al-2015-Book}
Y.~Bengio, I.~J. Goodfellow, and A.~Courville.
\newblock Deep learning.
\newblock Book in preparation for MIT Press, 2015.

\bibitem{bergstra2012random}
J.~Bergstra and Y.~Bengio.
\newblock Random search for hyper-parameter optimization.
\newblock {\em The Journal of Machine Learning Research}, 13(1):281--305, 2012.

\bibitem{bergstra2010theano}
J.~Bergstra, O.~Breuleux, F.~Bastien, P.~Lamblin, R.~Pascanu, G.~Desjardins,
  J.~Turian, D.~Warde-Farley, and Y.~Bengio.
\newblock Theano: A cpu and gpu math compiler in python.
\newblock In {\em Proc. 9th Python in Science Conf}, pages 1--7, 2010.

\bibitem{cadambi2010programmable}
S.~Cadambi, A.~Majumdar, M.~Becchi, S.~Chakradhar, and H.~P. Graf.
\newblock A programmable parallel accelerator for learning and classification.
\newblock In {\em Proceedings of the 19th international conference on Parallel
  architectures and compilation techniques}, pages 273--284. ACM, 2010.

\bibitem{chakradhar2010dynamically}
S.~Chakradhar, M.~Sankaradas, V.~Jakkula, and S.~Cadambi.
\newblock A dynamically configurable coprocessor for convolutional neural
  networks.
\newblock In {\em ACM SIGARCH Computer Architecture News}, volume~38, pages
  247--257. ACM, 2010.

\bibitem{chetlur2014cudnn}
S.~Chetlur, C.~Woolley, P.~Vandermersch, J.~Cohen, J.~Tran, B.~Catanzaro, and
  E.~Shelhamer.
\newblock cudnn: Efficient primitives for deep learning.
\newblock {\em arXiv preprint arXiv:1410.0759}, 2014.

\bibitem{cloutier1996vip}
J.~Cloutier, S.~Pigeon, F.~R. Boyer, E.~Cosatto, and P.~Y. Simard.
\newblock Vip: An fpga-based processor for image processing and neural
  networks.
\newblock In {\em microneuro}, page 330. IEEE, 1996.

\bibitem{coates2013deep}
A.~Coates, B.~Huval, T.~Wang, D.~Wu, B.~Catanzaro, and N.~Andrew.
\newblock Deep learning with cots hpc systems.
\newblock In {\em Proceedings of the 30th international conference on machine
  learning}, pages 1337--1345, 2013.

\bibitem{collobert2002torch}
R.~Collobert, S.~Bengio, and J.~Mari{\'e}thoz.
\newblock Torch: a modular machine learning software library.
\newblock Technical report, IDIAP, 2002.

\bibitem{cox1992ganglion}
C.~E. Cox and W.~E. Blanz.
\newblock Ganglion-a fast field-programmable gate array implementation of a
  connectionist classifier.
\newblock {\em Solid-State Circuits, IEEE Journal of}, 27(3):288--299, 1992.

\bibitem{czajkowski2012opencl}
T.~S. Czajkowski, U.~Aydonat, D.~Denisenko, J.~Freeman, M.~Kinsner, D.~Neto,
  J.~Wong, P.~Yiannacouras, and D.~P. Singh.
\newblock From opencl to high-performance hardware on fpgas.
\newblock In {\em Field Programmable Logic and Applications (FPL), 2012 22nd
  International Conference on}, pages 531--534. IEEE, 2012.

\bibitem{fang2011comprehensive}
J.~Fang, A.~L. Varbanescu, and H.~Sips.
\newblock A comprehensive performance comparison of cuda and opencl.
\newblock In {\em Parallel Processing (ICPP), 2011 International Conference
  on}, pages 216--225. IEEE, 2011.

\bibitem{farabet2009cnp}
C.~Farabet, C.~Poulet, J.~Y. Han, and Y.~LeCun.
\newblock Cnp: An fpga-based processor for convolutional networks.
\newblock In {\em Field Programmable Logic and Applications, 2009. FPL 2009.
  International Conference on}, pages 32--37. IEEE, 2009.

\bibitem{fowers2012performance}
J.~Fowers, G.~Brown, P.~Cooke, and G.~Stitt.
\newblock A performance and energy comparison of fpgas, gpus, and multicores
  for sliding-window applications.
\newblock In {\em Proceedings of the ACM/SIGDA international symposium on Field
  Programmable Gate Arrays}, pages 47--56. ACM, 2012.

\bibitem{hubel1968receptive}
D.~H. Hubel and T.~N. Wiesel.
\newblock Receptive fields and functional architecture of monkey striate
  cortex.
\newblock {\em The Journal of physiology}, 195(1):215--243, 1968.

\bibitem{jia2014caffe}
Y.~Jia, E.~Shelhamer, J.~Donahue, S.~Karayev, J.~Long, R.~Girshick,
  S.~Guadarrama, and T.~Darrell.
\newblock Caffe: Convolutional architecture for fast feature embedding.
\newblock {\em arXiv preprint arXiv:1408.5093}, 2014.

\bibitem{krizhevsky2014one}
A.~Krizhevsky.
\newblock One weird trick for parallelizing convolutional neural networks.
\newblock {\em CoRR}, abs/1404.5997, 2014.

\bibitem{krizhevsky2012imagenet}
A.~Krizhevsky, I.~Sutskever, and G.~E. Hinton.
\newblock Imagenet classification with deep convolutional neural networks.
\newblock In {\em Advances in neural information processing systems}, pages
  1097--1105, 2012.

\bibitem{lecun2015nature}
Y.~LeCun, Y.~Bengio, and G.~Hinton.
\newblock Deep learning.
\newblock {\em Nature}, 521:436--444, 2015.
\newblock Nature Publishing Group, a division of Macmillan Publishers Limited.
  All Rights Reserved.

\bibitem{ovtcharov2015accelerating}
K.~Ovtcharov, O.~Ruwase, J.-Y. Kim, J.~Fowers, K.~Strauss, and E.~S. Chung.
\newblock Accelerating deep convolutional neural networks using specialized
  hardware.
\newblock {\em Microsoft Research Whitepaper}, 2, 2015.

\bibitem{paul2006back}
K.~Paul and S.~Rajopadhye.
\newblock Back-propagation algorithm achieving 5 gops on the virtex-e.
\newblock In {\em FPGA Implementations of Neural Networks}, pages 137--165.
  Springer, 2006.

\bibitem{peemen2013memory}
M.~Peemen, A.~Setio, B.~Mesman, H.~Corporaal, et~al.
\newblock Memory-centric accelerator design for convolutional neural networks.
\newblock In {\em Computer Design (ICCD), 2013 IEEE 31st International
  Conference on}, pages 13--19. IEEE, 2013.

\bibitem{sankaradas2009massively}
M.~Sankaradas, V.~Jakkula, S.~Cadambi, S.~Chakradhar, I.~Durdanovic,
  E.~Cosatto, and H.~P. Graf.
\newblock A massively parallel coprocessor for convolutional neural networks.
\newblock In {\em Application-specific Systems, Architectures and Processors,
  2009. ASAP 2009. 20th IEEE International Conference on}, pages 53--60. IEEE,
  2009.

\bibitem{Schmidhuber}
J.~Schmidhuber.
\newblock Deep learning in neural networks: An overview.
\newblock {\em Neural Networks}, 61:85--117, 2015.
\newblock Published online 2014; based on TR arXiv:1404.7828 [cs.NE].

\bibitem{snoek2012practical}
J.~Snoek, H.~Larochelle, and R.~P. Adams.
\newblock Practical bayesian optimization of machine learning algorithms.
\newblock In {\em Advances in neural information processing systems}, pages
  2951--2959, 2012.

\bibitem{zhang2015optimizing}
C.~Zhang, P.~Li, G.~Sun, Y.~Guan, B.~Xiao, and J.~Cong.
\newblock Optimizing fpga-based accelerator design for deep convolutional
  neural networks.
\newblock In {\em Proceedings of the 2015 ACM/SIGDA International Symposium on
  Field-Programmable Gate Arrays}, pages 161--170. ACM, 2015.

\end{thebibliography}
%
\balancecolumns
\end{document}